\documentclass[final,5p,times,twocolumn]{elsarticle}

\usepackage{amssymb}
\usepackage{color}

\journal{}

\begin{document}

\begin{frontmatter}

\title{Background Measurements in the Gran Sasso Underground Laboratory}

\author{M.~Haffke} 
\author{L.~Baudis}
\author{T.~Bruch} 
\author{A.D.~Ferella} 
\author{T.~Marrod\'an~Undagoitia}
\author{M.~Schumann\corref{cor1}}
\cortext[cor1]{marc.schumann@gmx.net}
\author{Y.-F.~Te} 
\author{A.~van der Schaaf} 
\address
{Physik-Institut der Universit\"at Z\"urich\\
Winterthurerstrasse 190\\
8057 Z\"urich, Switzerland}

\begin{abstract}
The gamma background flux below~3000 keV in the Laboratori Nazionali del Gran Sasso (LNGS), Italy, has been measured  using a 3''~diameter NaI(Tl) detector at different underground positions: In hall A, hall B, the interferometer tunnel, and inside the Large Volume Detector (LVD). The integrated flux is 0.3--0.4~s$^{-1}$cm$^{-2}$ at the first three locations, and is lower by two orders of magnitude inside LVD. With the help of Monte Carlo simulations for every location, the contribution of the individual primordial isotopes to the background has been determined. Using an 11''~diameter NaI(Tl) detector, the background neutron flux in the LNGS interferometer tunnel has been estimated. Within the uncertainties, the result agrees with those from other neutron measurements in the main halls.
\end{abstract}


\end{frontmatter}


\section{\label{Introduction}Introduction}

Low background experiments studying rare processes -- such as neutrino experiments, or projects searching for dark matter or neutrinoless double beta decay -- are located at deep underground sites in order to reduce the background contribution from cosmic rays by several orders of magnitude. The experiments require precise knowledge of the background induced by gamma radiation and neutrons originating from their surrounding. These values are important to estimate the experiment's sensitivity and the expected signal to background level. 
 
To characterize an underground site the flux of gammas, neutrons, and muons (the \emph{external} background of the experiment, in contrast to the \emph{internal} background from the detector itself) are the most important parameters. 
Detailed Monte Carlo simulations of underground experiments use the spectra as input in order to predict background rates or to design appropriate shielding geometries. 

\begin{figure}[tb]
\begin{center}
\includegraphics[width=0.48\textwidth]{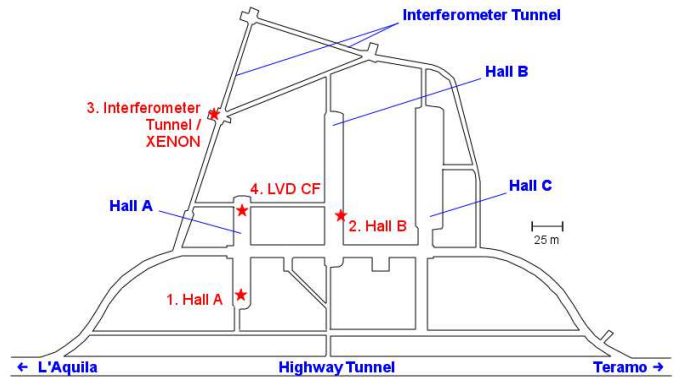}
\caption{Footprint of the LNGS underground laboratory. Gamma measurements were performed at locations 1--4 marked with a red star, a neutron flux estimation was done in the interferometer tunnel in front of XENON~(3).}
\label{fig::lngs}
\end{center}
\end{figure}

In this paper, we report on measurements of the gamma ray background at several locations in the Laboratori Nazionali del Gran Sasso (LNGS), Italy, which is one of the largest underground facilities worldwide at a depth of 3100 meters water equivalent (equivalent vertical depth, relative to a flat overburden). Only little information on the external background in the three large LNGS halls A, B, and C is available, and there is no published data on the background in the interferometer tunnel, where several experiments are located. Therefore we have measured the background in halls A and B, in the interferometer tunnel, and inside the LVD detector \cite{ref::LVD}. It was recently suggested that LVD could be used as a very efficient muon shield, effectively increasing the LNGS depth \cite{ref::arneodo2009}. Details on the locations of the measurements reported here are given in section \ref{sec::locations} and in Fig.~\ref{fig::lngs}.

The measurements were performed with NaI(Tl) detectors. These are very well suited for general gamma spectroscopy as they combine the easy operation of a scintillator crystal at room temperature with a high light output and good energy resolution. The latter is necessary to extract information about the background sources via the identification of characteristic gamma lines. Two different NaI(Tl) detectors were used for this study: A detector with 3''~diameter was used to measure the gamma background spectra as described in section \ref{sec::gamma}. The environment of each measurement was modeled with the GEANT4 toolkit \cite{ref::geant4}. By comparing the simulated spectra of the radioactive decay chains from $^{238}$U, $^{232}$Th, and $^{40}$K with the measured data, it is possible to deconvolute the measured spectrum into the individual contributions. This is a very powerful method to extract a maximum of information from the data. For example, a decay chain in secular equilibrium contains also alpha emitters, which cannot be seen in the gamma data, however, they might affect an experiment's neutron background via $(\alpha,n)$ reactions. Hence, the detailed breakdown into the different background is needed for realistic background simulations of existing or future experiments.
In section \ref{sec::neutron} we describe the estimation of the neutron flux in the interferometer tunnel using a large 11`` diameter NaI(Tl) detector.

\section{\label{sec::gamma}Gamma Flux Measurements}

The rock of the Gran Sasso mountain, the concrete used to stabilize the laboratory cavities, and all detector and supporting materials of any experimental setup contain primordial gamma emitting radioactive isotopes with very long half-lifes: $^{238}$U ($T_{1/2} =4.5 \times 10^9$~y), $^{232}$Th ($T_{1/2} =1.4 \times 10^{10}$~y), and $^{40}$K ($T_{1/2} = 1.3 \times 10^9$~y). A decay product of $^{238}$U, the noble gas $^{222}$Rn, also contributes to the natural background radioactivity in underground laboratories as it is emanated from the rock and -- being a noble gas -- can rather easily enter detector systems.
An isotope which is often present in the background of rare event searches is the cosmogenic $^{60}$Co ($T_{1/2} = 5.3$~y) which, however, was not observed in the measurements presented here.

Only little information on the gamma background of LNGS is published:
The integrated gamma flux in hall~C was measured to be $\Phi_C \sim 1 \ \textnormal{s}^{-1} \textnormal{cm}^{-2}$ \cite{ref::arpesella}, in hall~A the flux below 3000~keV was measured to be $\Phi_A \sim 0.73 \ \textnormal{s}^{-1} \textnormal{cm}^{-2}$ \cite{ref::bucci2009,ref::Bellini2010}. This paper presents new measurements of the gamma flux with a commercial 3''~NaI(Tl) detector. The small size of the detector allows measurements in a running laboratory, close to or even inside experimental setups.

\subsection{\label{DetectorDescription3inch}Experimental Method}

The detector\footnote{manufactured by {\it Saint Gobain Crystals and Detectors}} consists of a cylindrical NaI(Tl)  crystal with 7.62~cm diameter and 7.62~cm height. It is coupled to a 3''~diameter photomultiplier (PMT). Crystal and PMT are enclosed by an aluminum shell. The PMT signals are shaped using a spectroscopy amplifier and recorded with a 8k multichannel analyzer (MCA) with internal triggering.

The detector was set up to measure gamma energies up to 3000~keV. An external detection threshold of about 35~keV is given by the thickness of the Al-enclosure of the detector, however, gamma rays at these low energies constitute only a subdominant contribution to the background flux as they cannot escape the source because of the strong rise of the photoelectric effect's cross section. Energy calibration and energy resolution were obtained using standard calibrated gamma sources: $^{57}$Co, $^{60}$Co, $^{137}$Cs, $^{228}$Th, and $^{208}$Tl. The response of the NaI(Tl) scintillation detector is linear within 0.2\% over the energy range of interest. Its resolution $\Delta E$ (FWHM) is dominated by photo-electron statistics and can be described by $\Delta E/E = \sqrt{3.3~\textnormal{keV}/E}$; this relation was experimentally verified up to 2615~keV ($^{208}$Tl).

In order to determine the intrinsic detection efficiency of the setup, a Monte Carlo simulation using the precise detector geometry has been performed. From these simulations the ratio of events in the full absorption peak to events in the Compton continuum has been determined for monoenergetic gamma rays. Measurements with calibration sources of known activity were performed to determine the countrate in several peaks which was then compared to the expected value obtained from the simulation. The difference is due to the intrinsic detection efficiency.
Below 550~keV the efficiency was found to be $>0.99$ with a falling trend towards higher energies because of the finite size of the scintillator crystal. In agreement with the literature expectation for this type of detector \cite{ref::knoll99}, the efficiency is about 75\% at 1000~keV and 65\% at 2615~keV. The use of a small crystal is therefore no limiting factor in terms of efficiency.

\subsection{\label{sec::locations}Measurements}

Background measurements were performed at four underground locations at LNGS (cf.~Fig.~\ref{fig::lngs}). The gamma flux was measured in hall A (location 1), hall B (2), in the interferometer tunnel inside the XENON building (3), and in the the LVD Core Facility (LVD CF, 4) -- the innermost part of the Large Volume Detector \cite{ref::LVD,ref::arneodo2009}. The neutron flux was measured in the interferometer tunnel in front of XENON (3). The exact positions are described in this section and indicated in the layout of the LNGS underground facilities \cite{ref::lngs}, shown in Fig.~\ref{fig::lngs}. For each location, the measurement time was chosen such that the statistical uncertainty in the flux measurements was comparable to the systematic uncertainty from the efficiency determination.

1. Hall A: The smallest of the three main LNGS halls houses the experiments GERDA, LVD, CRESST, CUORE, and the ultra low background HPGe detector facility Gator (in the former GALLEX/GNO counting house). The NaI(Tl) detector was placed in front of the Gator house main entrance, directly on the concrete floor; data was taken for three days.

2. Hall B: This hall houses the large experiments ICARUS and WARP. The detector was placed between the WARP outer shield and the wall of the hall. Again, the gamma flux was measured for three days.

3. Interferometer Tunnel/XENON: The tunnels with smaller cross sections behind the main halls are connecting the large experimental areas and house the LNGS interferometer. Because of the increasing demand of underground space, the interferometer tunnel is nowadays also used for smaller experimental setups such as LArGe, LUNA, PULEX, and XENON and might be of interest for other experiments or underground R\&D setups. As no background data was available for the interferometer tunnel, we performed a measurement close to the XENON100 experiment \cite{ref::xe100} which is currently the dark matter experiment with the lowest background level. The gamma flux was measured for two days next to the XENON100 passive lead shield.

4. LVD Core Facility: The LVD Core Facility (LVD CF) is the innermost part of the \emph{Large Volume Detector} which is also located in hall~A. LVD is an observatory mainly devoted to the study of neutrinos from core collapse supernovae \cite{ref::LVD}. It has a highly modular structure, consisting of 1000~tons of liquid scintillator. The interest in the LVD CF was triggered by~\cite{ref::arneodo2009} who show that the muon flux in the CF, using LVD as a muon veto, is equivalent to a much deeper site. A flux of $\Phi_\mu < 3.6 \times 10^{-11}$~cm$^{-2}$s$^{-1}$ ($E_\mu>10$~MeV) untagged muons is estimated for the LVD CF, a value which is comparable to SNOLAB (6010~m.w.e.)\ with a predicted flux of $\Phi_\mu < 1.8 \times 10^{-11}$~cm$^{-2}$s$^{-1}$ \cite{ref::sudbury}. The steel/scintillator structure is also expected to lower the gamma background considerably, hence the 3"~NaI(Tl) detector was placed on top of the innermost LVD module, taking data for 10 days.

\begin{figure}[t!]
\begin{center}
\includegraphics[width=0.45\textwidth]{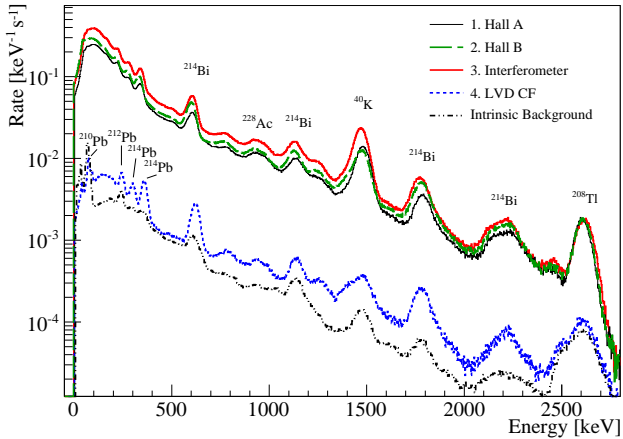}
\caption{Gamma spectra of LNGS hall A (location 1), hall B (2), the interferometer tunnel (inside the XENON building, 3), and the LVD Core Facility (4) measured with a 3''~NaI(Tl) detector. The LVD spectrum is much lower compared to the others because of the effective gamma shielding of the LVD steel/liquid scintillator structure. The intrinsic background spectrum of the detector is also shown.}
\label{compare_all}
\end{center}
\end{figure}

The gamma spectra measured at the four locations are shown in Fig.~\ref{compare_all}. While the flux in the halls of LNGS and the interferometer tunnel are similar, it is almost two orders of magnitude lower inside the LVD~CF due to the shielding provided by the LVD modules. The low energetic peaks in the LVD~CF spectrum are from lead isotopes in the LVD steel structure.

For a measurement of low gamma fluxes, it is essential to understand the intrinsic background of the detector itself. This has been measured underground at LNGS, inside a low background lead shield purged with high purity boil off nitrogen gas in order to suppress the background from $^{222}$Rn. The measuring time was 26 days; the result is also shown in Fig.~\ref{compare_all}. At the lowest energies, the intrinsic background spectrum is higher than the LVD~CF measurement. This is due to the 46.5~keV gamma line from $^{210}$Pb in the lead shielding.

\subsection{Analysis}

The setup used to measure the intrinsic background of the 3''~NaI(Tl) detector, including electronics and shielding, has been modeled in a dedicated Monte Carlo simulation. The contributions from the primordial isotopes $^{238}$U, $^{232}$Th, and $^{40}$K to the intrinsic background were determined by comparing the simulation with the measured spectrum. The scaling factors $f_c$ for each chain $c$ were treated as free parameters in a $\chi^2$-minimization fit. Taking into account the number of simulated events, the mass of the simulated material, and the measuring time, the individual activites $A_c$ can be calculated from $f_c$. The simulation shows satisfactory agreement with the data and leads to the following results: ($0.25\pm0.08$)~Bq/kg for $^{238}$U, ($0.8\pm0.2$)~Bq/kg for $^{232}$Th, and ($1.6\pm0.3$)~Bq/kg for $^{40}$K. 

Taking into account the intrinsic efficiency of the crystal, the integrated gamma flux below 3000~keV can be calculated from the measurements at the different locations. The results are summarized in Table~\ref{fluxes_gamma_act}, the errors consist of the statistical errors of the flux measurement, and the statistical and systematical error from the efficiency determination. The values for the halls~A and B and the interferometer tunnel are 2.5--3.5~times lower than the value quoted in \cite{ref::arpesella} for hall~C, and a factor 1.7--2.7~below the measurement of hall~A \cite{ref::bucci2009}. In both references, neither the exact position of the detector nor a detailed description of the measurement is given.
The intrinsic radioactivity of the detector had to be taken into account in the analysis of the low flux in the LVD~CF leading to a larger relative uncertainty; this contribution also causes the structures in the spectrum below 400~keV.

\begin{table}[tb]
\begin{center}
\caption{Gamma flux below 3000~keV, measured at several LNGS underground locations with a 3''~NaI(Tl) detector.}\label{fluxes_gamma_act}
\begin{tabular}{lcc}
\hline
\textbf{Location} & \textbf{Time} & \textbf{Flux [$\textnormal{s}^{-1}\textnormal{cm}^{-2}$]}    \\
\hline 
    1. Hall A   & 3 d & $(0.28\pm0.02)$  \\
    2. Hall B   & 3 d & $(0.33\pm0.03)$   \\
    3. Interferometer Tunnel & & \\
    \ \ \ (XENON building) & 2 d & $(0.42\pm0.06)$ \\
    4. LVD core facility & 10 d & $(0.005\pm0.001)$ \\ \hline \hline
\end{tabular}
\end{center}
\end{table}

A similar study performed with a larger NaI(Tl) detector in the LSM underground laboratory \cite{ref::ohsumi} showed that in underground laboratories the gamma ray flux below 4~MeV is dominated by radioactivity from the surrounding rocks. At higher energies, the intrinsic radioactivity of the NaI~crystal, neutron capture processes, and muon bremsstrahlung becomes relevant, however, the fluxes are suppressed by several orders of magnitude compared to the environmental gamma radioactivity. For the purpose of the gamma flux study presented here these contributions are completely negligible because of the low efficiency of the detector at these energies.

\begin{figure}[t!]
\begin{minipage}{8.5cm}
	\centering
	\includegraphics[width=1.\textwidth]{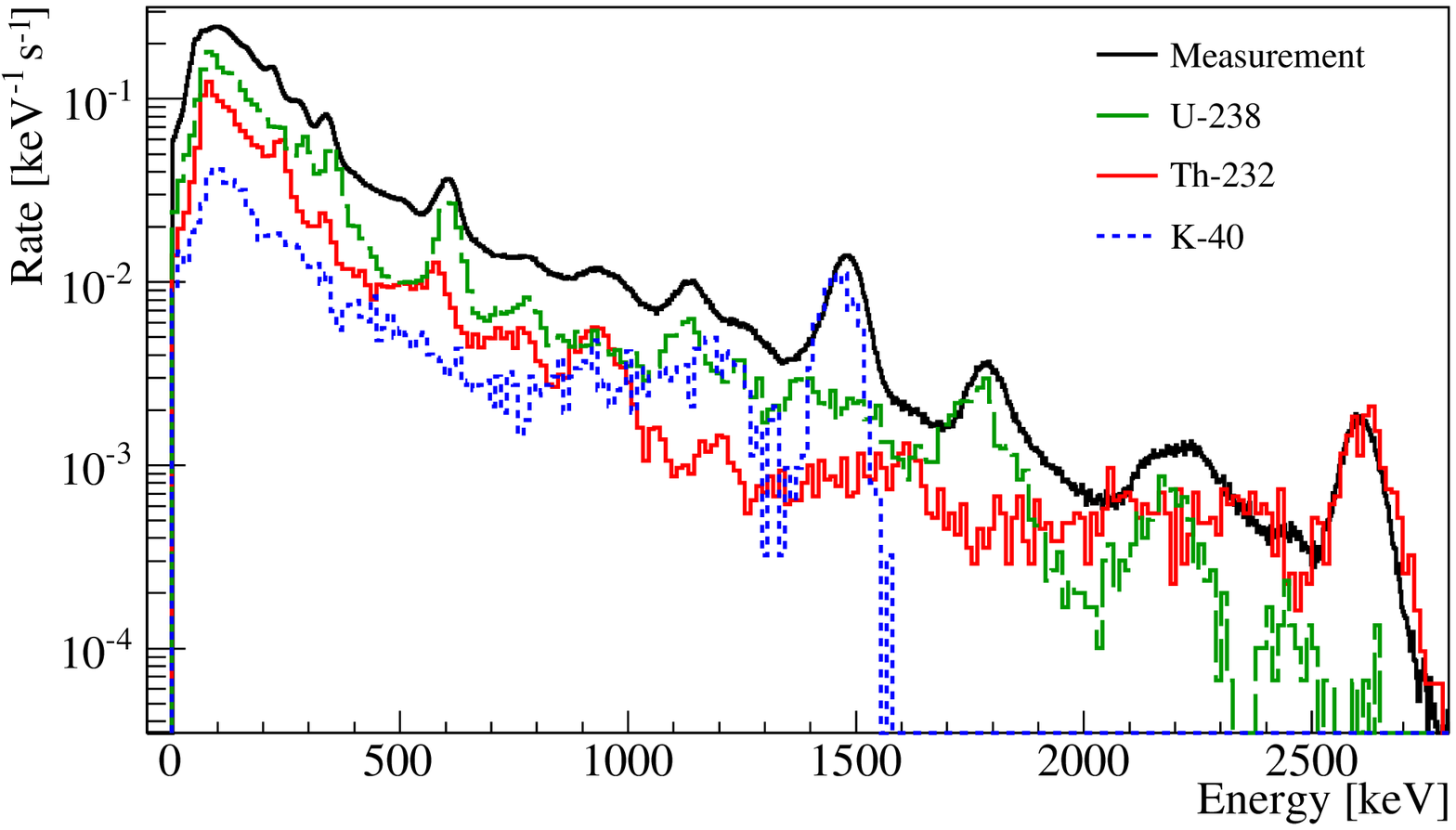}
\end{minipage}
\hfill
\begin{minipage}{8.5cm}
	\centering
	\includegraphics[width=1.\textwidth]{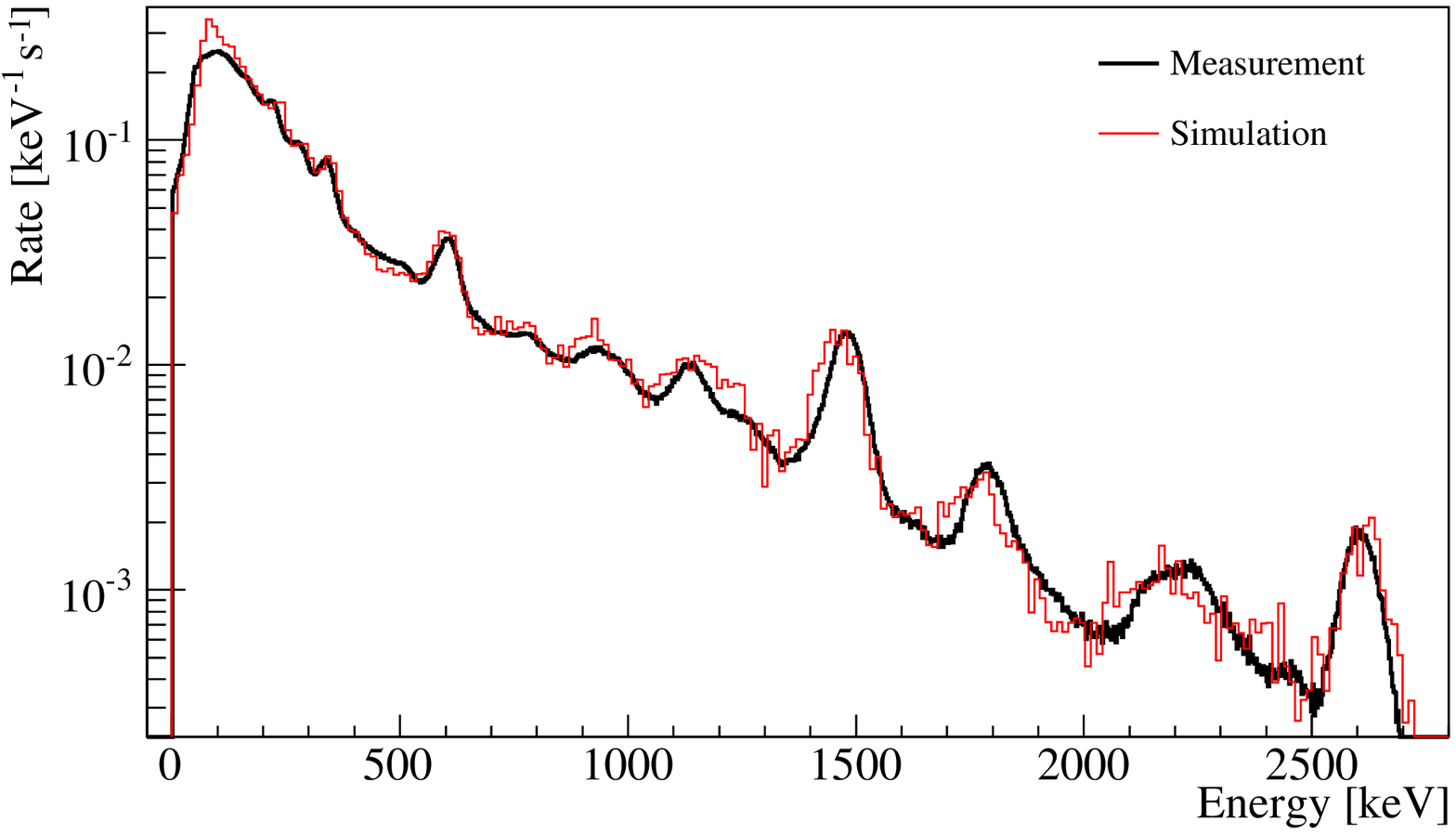}
\end{minipage}
\caption{Gamma spectrum of LNGS hall A (location 1). Top: Measured data and the individual contributions of $^{238}$U, $^{232}$Th, and $^{40}$K from of a Monte Carlo simulation of the setup. Bottom: Measured data and Monte Carlo sum spectrum agree very well over a very large energy range.}
\label{halla_Monte Carlo_vs_Daten}
\end{figure}

The contributions of the dominating isotopes ($^{238}$U, $^{232}$Th, and $^{40}$K) to the background were determined by comparing the measured spectra to the results of Monte Carlo simulations of the simplified measurement setup at every location. Besides the geometry, the simulation includes the measured energy resolution in order to yield realistic spectra. The efficiency is automatically taken into account by the simulation. The method works very well in the mid- and high-energy range, where most of the characteristic gamma-lines show up. It is somewhat limited at very low energies, where small (and possibly even unknown) details in the detector geometry start to play a role. Here, also the simplified geometry of the setup can affect the agreement, as smaller structural parts, which were not included in the simulation, might absorb low energetic gamma rays. The scaling factors $f_c$, however, are determined from the full measured spectrum and possible discrepancies at lowest energies do not affect the result.

For the halls~A and B and the interferometer tunnel the main contributor to the background radioactivity is the concrete used to support the underground cavities, while in the LVD Core Facility the radiation from the concrete is shielded by the LVD structure. Here, the radioactivity comes from the iron of the scintillator tanks, the PMTs of the detector, and the support structure. As an example, Fig.~\ref{halla_Monte Carlo_vs_Daten} (top) shows the individual contributions from the Monte Carlo simulation and the measured spectrum for hall~A (location 1). Fig.~\ref{halla_Monte Carlo_vs_Daten} (bottom) compares the Monte Carlo sum spectrum and the data: Simulation and real data agree very well except for the lowest energy region. The same holds for all other measured locations.

From the comparison of measurement and simulation, the background activity from the concrete in hall A, hall B, and the interferometer tunnel (XENON building), and the activity of the LVD materials can be determined. The results are summarized in Table~\ref{summary_gamma_act}, all numbers are given in Bq per kg of material. 

The concrete of the wall and the floor of the interferometer tunnel (close to XENON) were screened with the HPGe detector Gator \cite{ref::gator} in order to directly measure its radioactive contamination, cf.~Tab.~\ref{summary_gamma_act}. In particular the $^{40}$K contamination shows a large variation between the two samples, however, the numbers obtained for the concrete of the floor agree with the results from the NaI(Tl) measurement. Due to background contributions other than the concrete, all NaI(Tl) values are slightly higher.

\begin{table}[h!]
\begin{center}
\begin{minipage}{\linewidth} 
\renewcommand{\footnoterule}{}
\caption{Gamma activities of the primordial isotopes (in Bq/kg) as determined from measurements with a 3''~NaI(Tl) detector. \label{summary_gamma_act} }
\begin{tabular}{lcccc}
\hline \hline
\textbf{Location} & \textbf{$^{238}$U}    & \textbf{$^{232}$Th}  & \textbf{$^{40}$K} \\
\hline 
    1. Hall A   & $11.7\pm3.9$   & $14.8\pm2.8$ & $62\pm14$\\
    2. Hall B   &  $19.6\pm4.9$   & $13.2\pm2.7$  & $52\pm10$\\
    3. Interferometer & $37.8\pm7.3$  &  $10.9\pm2.8$ & $206\pm37$\\
    4. LVD CF  & $1.2\pm0.4$  & $0.34\pm0.07$ & $1.04\pm0.32$\\ \hline 
    Concrete (Floor)\footnote{These concrete samples taken from the interferometer tunnel were screened in a HPGe detector in order to directly measure the radioactive contamination. They are to be compared to the NaI(Tl) results for location {\it 3.~Interferometer}.} & $26 \pm 5$ & $8 \pm 2$ & $170 \pm 27$\\
    Concrete (Wall)$^a$ & $15 \pm 2$ & $3.8 \pm 0.8$ & $42 \pm 6$ \\
\hline \hline
\end{tabular}
\end{minipage}
\end{center}
\end{table}

\section{\label{sec::neutron}Neutron Flux Estimate for the Interferometer Tunnel}

Neutrons constitute the most critical background for many experiments searching for rare processes as they are harder to shield than gamma rays. In some cases they cannot be distinguished from signals, for example in experiments searching for nuclear recoils from WIMP dark matter. The dominant sources of neutrons in underground laboratories are $(\alpha,n)$ reactions on light elements (C, O, F, Na, Mg, Al, Si), where the $\alpha$-particle comes from the primoridal decay chains, and spontaneous fission, mainly of $^{238}$U \cite{ref::wulandari2004}. 
Because of the depth, the flux of high energetic neutrons ($>10$~MeV) induced by cosmic ray muons is 2--3~orders of magnitudes lower \cite{ref::Araujo2005}. The neutron flux from 0--10~MeV in hall~A of LNGS was found to be $\Phi_{n,A}= (3.81 \pm0.11)\times 10^{-6} \ \textnormal{s}^{-1} \textnormal{cm}^{-2}$ (value from \cite{ref::wulandari2004}, using measurements from \cite{ref::belli1989} and \cite{ref::arneodo1999});
for an overview on neutron background measurements in the LNGS hall~A, cf.~\cite{ref::wulandari2004}. 

In this section, we present an estimation of the neutron flux in the interferometer tunnel (in front of the XENON building, location~3) using a 11''~NaI(Tl) detector. The application of this type of detector for neutron flux measurements has been reported before \cite{ref::ohsumi}; it exploits the capture of thermal neutrons on iodine:
$$
n + ^{127}\textnormal{I} \rightarrow ^{128}\textnormal{I} + \gamma(6.826~\textnormal{MeV}), \qquad \sigma = 6.15 \ \textnormal{b}.  
$$
The reduction of environmental gamma background in the energy region around 6.8~MeV allows to derive the integral neutron flux with this detectors by measuring the area of the gamma peak from the capture reaction. 
Compared to other easily available detectors, using this large NaI(Tl) crystal has the advantage of a sizeable area coverage and a massive active target in a compact geometry, which facilitates studies of the intrinsic background. At the same time, the NaI(Tl) detector suffers from a rather low absorpton cross secton leading to a low efficiency, and some contamination from the gamma background. Therefore, this
kind of measurement cannot compete with efforts using dedicated neutron counters \cite{ref::belli1989} such as $^3$He or BF$_3$. In the absence of other data, however, it still provides important information on the overall neutron background level and can be used for cross checks with measurements performed at other locations.

\subsection{\label{sec::DetectorDescription11inch}Experimental Method}

The NaI(Tl)~crystal has a diameter of 27.7~cm and a height of 33~cm. Seven 3''~PMTs are coupled to one side of the crystal, however, only 6 PMTs were operational during this measurement. The detector is enclosed in aluminum and the crystal is additionally surrounded by 5~cm of polyethylene for thermal insulation and neutron moderation. 
The analog signals of the 6~PMTs are added using a linear FAN-IN, shaped with a spectroscopy amplifier, and recorded using a 8k~MCA. 

Energy calibration and resolution have been determined with standard gamma calibration sources. The resolution, measured at energies up to 2.6~MeV, is 9\% at 2~MeV and extrapolated to be 5\% at 7~MeV. The energy threshold for the neutron measurements was then raised to about 2~MeV, still well below the region of interest for neutron capture.

The neutron detection efficiency was determined with a calibrated AmBe neutron source, leading to $\epsilon = (5.3 \pm 0.2) \%$. The spectrum of the AmBe source covers the same energy range (0--10~MeV) as the expected underground neutron spectrum (cf.~\cite{ref::wulandari2004}) with a slightly higher mean energy. The shape of the expected underground spectrum is also steeper. (The flux of muon induced neutrons is several orders of magnitude suppressed and can be neglected here.) Therefore, the calibrations with AmBe might lead to a slight overestimation of the derived underground neutron flux.

\subsection{\label{NMeasurement}Measurement and Analysis}

\begin{figure}[tb]
\begin{center}
\includegraphics[width=0.45\textwidth]{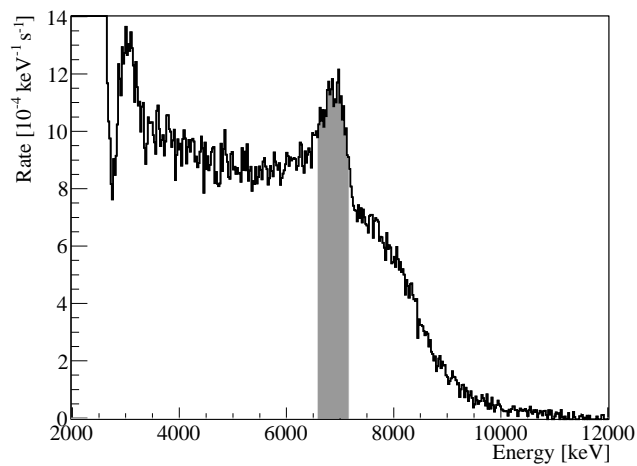}
\caption{Neutron response spectrum of an 11''~NaI(Tl) crystal to an AmBe source. The intrinsic background of the crystal and the high energetic muon background have been subtracted. The neutron capture peak is clearly visible, the region used for the analysis was determined from this measurement and is marked in gray.}\label{fig::pure_neutron}
\end{center}
\end{figure}

Neutrons capture on $^{127}$I does not only lead to a peak at 6.8~MeV, but also to events at lower energies when only a fraction of the full gamma energy is absorbed in the crystal. Elastic neutron scattering is only relevant at much lower energies. Events from 7--10~MeV are caused by gamma rays from neutron capture processes in other materials than the NaI crystal \cite{ref::ohsumi}. Entries at much higher energies are from muons. In order to extract a pure neutron response spectrum, the intrinsic background of the 11''~NaI(Tl) detector has been measured in a lead shield above ground at the Paul Scherrer Institut (PSI, Switzerland) for 34~days. It has been compared with the response to a continuous neutron spectrum from an AmBe source, also measured with the detector inside the lead shield. The difference spectrum is shown in Fig.~\ref{fig::pure_neutron}. The 6.8~MeV peak from neutron capture is clearly visible on top of the continuous spectrum that extends up to 10~MeV. Based on this measurement, a region around the peak was defined as the region of interest for the analysis.

With a NaI(Tl) detector it is not possible to clearly separate the signal from neutron capture on~I from other high energetic interactions. When no background is subtracted, the inferred neutron flux is therefore always an upper limit on the real flux, since some of the entries treated as neutrons are from external gamma rays. 
An estimate of the continuous spectrum below the peak is necessary in order to also determine a lower limit, and an AmBe calibration with and without lead shield (to block high energetic gammas from outside) was performed. This measurement was realized above ground and also contains a muonic component, leading to an overestimation of the gamma rate coming from outside and hence an underestimation of the neutron flux. 

\begin{figure}[tb]
\begin{center}
\includegraphics[width=0.45\textwidth]{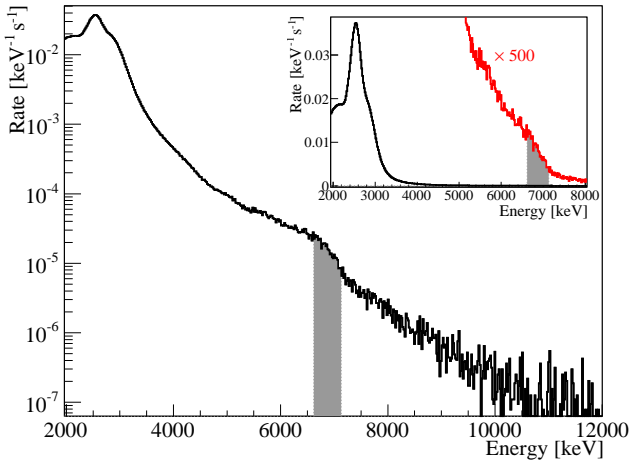}
\caption{Background spectrum measured in the LNGS interferometer tunnel with a 11''~NaI(Tl) detector. The gray region shows the location of the expected peak from thermal neutron capture on~I, which can be used to give limits on the integral neutron flux. For comparison with Fig.~\ref{fig::pure_neutron}, the inset figure shows the same spectrum with a linear scale (also with $\times 500$ magnification).}
\label{fig::nspec}
\end{center}
\end{figure}

In order to estimate the neutron flux in the LNGS interferometer tunnel, in front of the XENON building (location 3), the high energetic gamma flux was measured for 37~days. In comparison to the AmBe calibrations, the 6.8~MeV peak from neutron capture is hardly visible due to the very low neutron rate and the overlap with high energetic gamma events from reactions outside the detector. The measured spectrum and the region of interest are shown in Fig.~\ref{fig::nspec}.

Taking into account the width of the region of interest and the detection efficiency, a neutron flux of 
$$
1.5 \times 10^{-6} \ \textnormal{s}^{-1} \textnormal{cm}^{-2} < \Phi_{n,\textnormal{\footnotesize interferometer}} < 4.6 \times 10^{-6} \ \textnormal{s}^{-1}\textnormal{cm}^{-2}
$$
could be determined for the interferometer tunnel. Because of the neutron spectrum which is steeply falling with energy, and because the capture cross section is higher for lower neutron energies, this number is dominated by the thermal neutron flux. 

The upper and the lower limit are not defined in a statistical way, but should be considered as absolute upper and lower limits from a systematic point of view as the neutron peak cannot be fully separated from background: For the upper limit, the intrinsic background spectrum has not been taken into account, leading to an overestimation of the neutron flux. In case of the lower limit, the intrinsic background contaminated with a small component from atmospheric muons has been subtracted, leading to an underestimation of the neutron flux. The result is in good agreement with the values summarized in \cite{ref::wulandari2004} for the large LNGS halls -- there is no indication of a considerably increased or decreased neutron background in the interferometer tunnel.

\section{\label{sec::summary}Summary}

The precise knowledge of the radioactive gamma background in an underground laboratory is essential to design experiments searching for rare events. In case of the Laboratori Nazionali del Gran Sasso (LNGS), one of the largest underground facilities worldwide, a small number of neutron background measurements have been performed, mainly in hall~A \cite{ref::wulandari2004}. Numbers for the gamma flux are only available for halls~A \cite{ref::bucci2009} and~C \cite{ref::arpesella}.

We used NaI(Tl) scintillation detectors to precisely measure the gamma flux at four different LNGS locations (hall~A, hall~B, interferometer tunnel/XENON building, and the LVD core facility \cite{ref::arneodo2009}), and to estimate the integral neutron flux in the interferometer tunnel. In case of the gamma flux -- which is dominated by the concrete covering the walls and the floor of the laboratory -- we compared the measurements with detailed Monte Carlo simulations in order to derive the individual contributions of the primordial isotopes $^{238}$U, $^{232}$Th, and $^{40}$K. This information can be used as an important input parameter for shield and background studies for existing or next generation rare event search experiments, such as ton-scale dark matter search experiments.

All integral gamma fluxes are a bit lower than the published values. The values obtained for the interferometer tunnel agree with the intrinsic radioactivity of the concrete of the same location, measured with a HPGe spectrometer. 

The neutron flux estimated in the interferometer tunnel agrees with the fluxes obtained in the large LNGS halls.

\section*{Acknowledgments}

This work has been supported by the Universit\"at Z\"urich, by the Swiss National Foundation (grant numbers 118119 and 126938), and by the Volkswagen Foundation.  We would like to thank F.~Arneodo and S.~Fattori for the support during the measurements at LNGS, W.~Bertl for the support at PSI, and A.~Kish for his help with some simulations.


\end{document}